\documentclass{aa}
\usepackage{graphics,aabib99}

\def \kms          {\hbox{km$\,$s$^{-1}$}}
\def \ergpers      {\hbox{ergs$\,$s$^{-1}$}}
\def \cubiccm	   {cm$^{-3}$}
\def \lesssim      {\mathrel{\hbox{\rlap{\hbox{\lower4pt\hbox{$\sim$}}}\hbox{$<$}}}}

\makeatletter
\def\citeonlyyear#1{{\def\@cite##1##2{{##1\if@tempswa , ##2\fi}}\cite*{#1}}}
\makeatother
\begin{document}

   \thesaurus{     		% 
              (09.02.1;  	% ISM: bubbles,
               09.07.1;  	% ISM: general,
               09.19.1;  	% ISM: structure,
               11.07.1;  	% Galaxies: general,
               99.09.1 NGC 5775;% Galaxies: NGC 5775,
               11.19.2)} 	% Galaxies: spiral
   \title{NGC~5775: anatomy of a disk-halo interface}

   \subtitle{}

   \author{S.-W. Lee\inst{1}
          \and
          Judith A. Irwin\inst{2}
	  \and 
	  R.-J. Dettmar\inst{3}
	  \and
	  C. T. Cunningham\inst{4}
	  \and
	  G. Golla\inst{3}\fnmsep\thanks{\emph{Present address:} Gesellschaft
          f\"ur Unix- und Netzwerkadministration, R\"aterstr. 26,
          D-85551 Kirchheim, Germany; email: golla@genua.de}
          \and
          Q. D. Wang\inst{5}
          }

   \offprints{S. -W. Lee}

   \institute{Astronomy Department, University of Toronto, 
		60 St. George Street, Toronto, Ontario, Canada M5S
		3H8\\
		email: swlee@astro.utoronto.ca
         \and
             Department of Physics, Queen's University, Kingston, 
	     Ontario, Canada K7L 3N6\\
             email: irwin@astro.queensu.ca
	 \and
	     Astronomisches Institut der Ruhr-Universit$\ddot{a}$t Bochum,
	     Universit$\ddot{a}$tsstr. 150/NA 7, D-44780 Bochum,
              F. R. Germany \\
		email: dettmar@astro.ruhr-uni-bochum.de
	 \and
	     National Research Council of Canada, Herzberg Institute
		of Astrophysics, 5071 West Saanich Road, Victoria, 
		BC, Canada V9E 2E7 \\
		email: charles.cunningham@nrc.ca
	 \and
             Department of Astronomy, University of Massachusetts, 
	     B-524 Lederle Graduate Research Tower, Amherst, MA,
	     01003, USA \\
		email: wqd@astro.umass.edu
             }

   \date{Received <date>; accepted <date>}

   \maketitle

   \begin{abstract}

We present the first high-resolution study of the disk-halo interface
in an edge-on galaxy (NGC~5775) in which every component of the
interstellar medium is represented and resolved (though not all to the
same resolution).  New single-dish CO J=2-1 and CO J=1-0 data, ROSAT
X-ray data, and HIRES IRAS data are presented along with HI data which
emphasizes the high latitude features.  In conjunction with previously
published radio continuum (6 and 20 cm) and H$\alpha$ data, we find
spatial correlations between various ISM components in that all
components of the ISM are present in the disk-halo features (except
for CO for which there is insufficient spatial coverage).  The HI
features extend to $\sim$ 7 kpc above the plane, form loops in
position-position space, in one case, form a loop in position-velocity
space, and are also observed over a large velocity range.  This
implies that the disk-halo features represent expanding
supershells. However, the shells may be incomplete and partially
open-topped, suggesting that we are observing the breakup of the
supershells as they traverse the disk-halo interface.  There is some
evidence for acceleration with {\it z} and both redshifted and
blueshifted velocities are present, although the gas which is lagging
with respect to galactic rotation dominates.  The radio continuum
spectral index is flatter around the shell rims and we show that this
cannot be due to a contribution from thermal gas but rather is due to
intrinsic flattening of the non-thermal spectral index, suggesting
that shocks may be important in these regions.  The H$\alpha$ emission
is located {\it interior} to the HI.  For feature F3, the H$\alpha$
emission forms the interior ``skin" of the HI shell, yet there appears
to be a minimum of in-disk star formation immediately below the
feature.  We present a picture of a ``typical" HI supershell which
accelerates and breaks up through the disk-halo interface.  Such a
feature is likely internally generated via an energetic event in the
disk.

      \keywords{ISM: bubbles -- ISM: general -- ISM: structure -- 
		Galaxies: general -- Galaxies: NGC 5775 -- Galaxies: 
		spiral}
   \end{abstract}

%
%________________________________________________________________

\section{Introduction}

Studies of external edge-on galaxies provide an excellent bird's eye
view of galaxy halos and the disk-halo interface without the confusion
or distance ambiguities which are present in our own Milky Way.  Such
observations require high resolution and good signal-to-noise (S/N).

When a variety of galaxies is considered, disk-halo activity has been
observed in all components of the interstellar medium (ISM), including
molecular gas traced by CO (e.g. Irwin \& Sofue
\citeonlyyear{irwin:96}), dust as observed in the infra-red (e.g. Koo
et al. \citeonlyyear{koo:91}) and optical (e.g., Howk \& Savage
\citeonlyyear{howk}), neutral hydrogen (e.g. Irwin \& Seaquist
\citeonlyyear{irwin:90}), H$\alpha$ emitting 10$^4$ K ionized gas
(e.g. Pildis et al. \citeonlyyear{pildis}), the radio continuum
emitting cosmic ray component (e.g.  Hummel et
al. \citeonlyyear{hummel}) and the X-ray emitting hot coronal gas
(e.g.  Bregman \& Pildis \citeonlyyear{bregman:94}; Wang et
al. \citeonlyyear{wang}; Dahlem et al. \citeonlyyear{dahlem:96}) -- as
well as many other examples.  However, little has been done in the way
of systematic multiwavelength studies to see how the components relate
to each other, to discrete star forming regions in the disk, and to
the global properties of the galaxy.  To this end, we have undertaken
a multi-wavelength study of the edge-on, IR-bright galaxy, NGC~5775,
and present here the first comparative study of the disk-halo
interface in another galaxy in which every component of the
interstellar medium is represented.

NGC~5775 ({\it D} = 24.8 Mpc, $H_\mathrm{0}$ = 75 km s$^{-1}$
Mpc$^{-1}$) is an edge-on ({\it i} = 86\degr) galaxy with a high star
formation rate ($L_\mathrm{FIR}$ = 2.6 $\times$ 10$^{10}$ $L\sun$). An
optical image of the galaxy is shown in Fig.~1a.
NGC~5775 is interacting with its more face-on companion, NGC~5774,
4\farcm3 to the north-west (NW) from which it appears to be receiving
a supply of HI \cite{irwin:94a}, and possibly also with a smaller
galaxy, IC~1070, 3\farcm9 to the south-west (SW).  Duric et
al. \cite*{duric:98} have modeled the global radio continuum halo of
NGC~5775 in terms of two components: a broad scale steep spectrum halo
with 1/e scale height of 3.5 kpc, and a smaller flatter spectrum
component of $\sim$ 1 kpc.  Dettmar \cite*{dettmar:92} made a
preliminary comparison of the H$\alpha$ and radio continuum emission
in NGC~5775 and pointed out the spatial coincidence of the H$\alpha$
filamentary features and the radio continuum spurs.  Collins et
al. \cite*{collins} found the diffuse ionized gas (DIG) layer in
NGC~5775 to have a scale height of 800-900~pc in regions away from any
obvious discrete high-{\it z} features.  They made a comparison of
their H$\alpha$ images with both the radio continuum and total
intensity HI, finding a good general correlation in the high-{\it z}
features in all three components. More recently, an optical emission
line study by Rand \cite*{rand:00} shows a rise in the [O
III]/H$\alpha$ line ratio with {\it z}, indicating that
photoionization at a single temperature cannot alone be producing the
emission; shocks and/or a change in temperature is required.  A
decrease in rotational velocity with {\it z} is also observed, with
the highest latitude gas showing essentially no rotation.

In this paper, we present new CO J=1-0 and CO J=2-1 observations, new
HIRES infra-red images, and a new X-ray image of NGC~5775.  We also
present new results on the HI distribution using the same data as in
Irwin \cite*{irwin:94a} but dealing more specifically with the high
latitude gas.  In addition, we present an overview of previous results
from other wavebands, highlighting features which may be related to
the interaction, the in-disk distribution, and the high latitude and
halo emission.  Since the data are voluminous, in this paper, we focus
primarily on spatial correlations of the different ISM components.  We
also pay special attention to the high-{\it z} features and their
relationship with the distributions in the disk.  As the data are not
all to the same resolution, we compare components which are most
similar in resolution to each other.

%__________________________________________________________________

\section{Observations and Data Reduction}

\subsection{Radio Continuum}

The radio continuum observations are taken from Duric et
al. \cite*{duric:98} and were obtained with the Very Large Array (VLA)
over several years at both 6 and 20 cm wavelengths in several
configurations of the array.  Selected images, which have been cleaned
and self-calibrated, are shown in Fig.~1 and
Fig.~2, i.e., a 20 cm B + C + D Array 20 cm image at
23\arcsec\ resolution (Fig.~1b), a 6 cm C + D array
image at 15\arcsec\ resolution (Fig.~2a), a 20 cm B + C +
D array image at 15\arcsec\ resolution (Fig.~2b) and a 20
cm B array image at 5\arcsec\ resolution (Fig.~2d). The
spectral index map made from the 6 and 20 cm maps at 15\arcsec\
resolution is also presented in Fig.~2c. The rms noise
levels are indicated in the figure caption.

\subsection{Neutral Hydrogen}

Neutral hydrogen data are taken from Irwin \cite*{irwin:94a} and were
obtained using the VLA in its B/C configuration with a velocity
resolution of 41.7 km s$^{-1}$.  Two (Right Ascension - Declination -
Velocity) cubes were made, representing naturally and uniformly
weighted data, with resulting rms noise per channel of 0.4 and 0.5 mJy
beam$^{-1}$, respectively.  The integrated intensity (zeroth moment)
maps made from these data are presented in Fig.~1c
(natural weight) and Fig.~1d (uniform weight).  The
most obvious HI extensions are labelled F1 to F3 in
Fig.~1d.

\subsection{Far Infra-red (FIR)}

Four HI-RES infra-red images (at 12, 25, 60 and 100 $\mu$m) were
obtained by request to the Infra-red Processing and Analysis Center
(IPAC).  The HI-RES images were obtained from the Infra-Red
Astronomical Satellite (IRAS) observations using the Maximum
Correlation Method as described in Aumann et al. \cite*{aumann}.  This
algorithm makes use of the fact that the spatial separation between
flux measurements (``footprints") is typically less than the full
width half maximum (FWHM) of any individual detector, i.e. the sources
have been oversampled.  The reconstructed beam FWHMs constitute
improvements by factors of 3.8, 3.7, 3.5, and 2.6 over the mean FWHMs
of the individual detectors for the 12, 25, 60, and 100 $\mu$m bands,
respectively.  The HI-RES maps, the result of 20 iterations of the
algorithm, are shown in Fig.~3a to d.

\subsection{H$\alpha$}

The H$\alpha$ image was obtained from the European Southern
Observatory (ESO) 3.5 m New Technology Telescope (NTT).  The
integration time was 60 minutes in a filter of FWHM = 75~\AA\ and the
seeing was 1\farcs1. The H$\alpha$ image was obtained after R-band
subtraction and is shown in Fig.~4a.  Calibration was
done using an R-band magnitude of NGC~5774 and the level of the rms
noise of the map is
6.6$\times$10$^{-18}$~\ergpers~cm$^{-2}$~arcsec$^{-2}$.  Thus, this
image is of comparable sensitivity to that of Collins et
al. \cite*{collins}.  There is also a background gradient at about the
level of the noise. This image, without the quantitative scale, has
previously been presented in Dettmar \cite*{dettmar:92}.

\subsection{CO}

The molecular gas in the central region of the galaxy was mapped in
the CO J=1-0 line using the 15 m Swedish-ESO Submillimetre Telescope
(SEST)\footnote{The SEST is operated jointly by ESO and the Swedish
National Facility for Radio Astronomy, Onsala Space Observatory,
Chalmers University of Technology.} during July 27-31 1990 and in the
CO J=2-1 line using the James Clerk Maxwell Telescope
(JCMT)\footnote{The JCMT is operated by the Joint Astronomy Centre in
Hilo, Hawaii on behalf of the parent organizations Particle Physics
and Astronomy Research Council in the United Kingdom, the National
Research Council of Canada and The Netherlands Organization for
Scientific Research.} during the period from July 1993 to February
1995. The half-power-beam-width (HPBW) of the SEST at 115~GHz is
43\arcsec\ and that of the JCMT at 230~GHz is 21\arcsec. Pointing
accuracies are roughly 5\arcsec\ at both telescopes. At the JCMT, the
absolute flux calibration was checked against sources of known flux
during each observing run and was found to be good to within 20\%.
  Spectra at 19 points for CO J=1-0 and 36 points for CO J=2-1
were integrated over velocity and interpolated onto a map grid.  The
resulting maps are shown in Fig.~4b and
Fig.~4c.  Full details are presented in Lee
\cite*{lee}. The results of a line ratio analysis will be presented in
a separate paper.

\subsection{X-ray}

X-ray observations were obtained from the ROSAT satellite (sequence
number=RP800478N00) between July 29 and August 13, 1993. The total
exposure time was 6140 seconds and the angular resolution is about
30\arcsec.  An overlay of these results on an optical image is shown
in Fig.~5.  The X-ray spectrum is characterized by an
optically-thin thermal plasma with average temperature of 0.4 keV
(about 5 million K). The limited spatial resolution and counting
statistics of the X-ray data prevent us from a spatially-resolved
spectral analysis. Thus a significant contribution of point-like
sources, which typically have relatively hard X-ray spectra, in the
galactic disk region may be important.

\section{Results}
\subsection{The Interaction}
\label{the_interaction}

\subsubsection{Tidal Bridges}

The environment of NGC~5775 is illustrated in Fig.~1.
The interaction of NGC~5775 with its companion, NGC~5774, is indicated
by the presence of two bridges.  The bridges are especially apparent
in HI (Fig.~1c and d) and the HI velocity field
suggests mass transfer from NGC~5774 to NGC~5775 \cite{irwin:94a}.
There is no strong optical disturbance in either galaxy 
 (Fig.~1a), but faint emission
is seen along both bridges on the Palomar Observatory Sky Survey
(POSS) blue print indicating that a stellar component is indeed
present.  Although not shown here, our H$\alpha$ image shows emission
along the southern bridge (also confirmed by Collins et
al. \citeonlyyear{collins}). If the ionization is due to hot young
stars, then star formation must be occurring in the bridge itself
since the lifetimes of such stars are shorter than the interaction
timescale (see Irwin \& Caron \citeonlyyear{irwin:94b}).  Another
tracer of hot young stars, non-thermal radio continuum emission, also
occurs along the southern bridge (Fig.~1b).  Note that,
when the resolution is sufficient, considerable structure is observed
in the bridges.

An overlay of radio continuum (from Fig.~1b, rotated)
as well as the HIRES 12~$\mu$m and 25~$\mu$m (from Fig.~3a
and b) on the HI image (from Fig.~1c) is shown in
Fig.~6.  Particularly interesting is the FIR
emission between the galaxies which, at 12~$\mu$m and 25~$\mu$m, occurs
only along the southern bridge.  At 12~$\mu$m
(Fig.~6b), the emission is seen as a two-pronged
spur extending from NGC~5775 with the southern prong the strongest.
At 25~$\mu$m (Fig.~6c), a complete bridge connects
the galaxies.  This bridge starts as a strong spur from NGC~5775 (at
RA = 14$^h$53$^m$51$^s$, $\delta$ = 03\degr33\arcmin49\arcsec) which
corresponds to a radio continuum spur (Fig.~6a,
third contour).  However, the FIR bridge is offset $\approx$
30\arcsec\ to the NE (i.e. towards the inside) of the centroid of the
HI / radio continuum bridge.  Together with
the two-pronged spur at 12~$\mu$m, this suggests that the dust could
be in a sheath around the radio continuum emission.  Temperature
gradients and/or different dust properties may also be contributing to
the different morphologies seen between the IR wavebands.  In lower
resolution 60~$\mu$m and 100~$\mu$m images (Fig.~3), the two
galaxies appear enveloped in emission.

The x-ray emission (Fig.~5)
shows two bright peaks north and south of the companion galaxy, NGC~5774,
and another weaker peak to the east of it. The bright emission
peaks are most likely background active galactic nuclei or foreground
stars. The fainter peak occurs within the northern bridge, at RA =
14$^h$53$^m$50\fs7, $\delta$ = 03\degr35\arcmin10\arcsec, and is near,
but not exactly coincident with a decrease in the HI emission. If this
peak is caused by HI absorption of the x-ray emission in the bridge,
this would place NGC~5774 on the near side of the hot gas.  This is
consistent with the fact that NGC~5774 is blueshifted with respect to
NGC~5775 and may be in front of it.
 
It would appear that the interaction is causing various components in
NGC~5774, including both stars and the ISM, to be transferred to
NGC~5775 through two bridges.  There is evidence for star formation in
the southern, but not the northern, bridge.  Thus the interaction has
either been sufficiently disruptive that it has induced star formation
in the bridge, or it has been sufficiently quiescent that star
forming regions previously present in NGC~5774 may simply have
continued their activity through the bridge as they transfer to
NGC~5775.  Given the continuity of the ISM in NGC~5774 into the
bridges and the lack of any strong optical disturbance in the
galaxies, the latter may be the case.  Follow-up observations of the
ionized regions in the bridge would be useful to confirm whether they
indeed trace star forming regions, or whether shocks and collisionally
ionized gas are present.

\subsubsection{South-East Feature (Tidal Tail?)}
\label{southeasttidal}

At the far south-east tip of NGC~5775, there is a feature which may be
tidal in origin.  It is most obvious in the H$\alpha$ map
(Fig.~4a) at $\alpha$ = 14$^h$54$^m$03, $\delta$ =
03\degr31\arcmin30\arcsec, where an arc-like string of HII regions
extends abruptly away from the end of the disk.  The H$\alpha$
emission is located along the northern ridge of a radio continuum
extension in roughly the same direction (Fig.~2b and d).
The feature can also be seen as a small extension in neutral hydrogen
(Fig.~1c and d) and as a south-east `knob' of FIR
emission on the 25~$\mu$m map (Fig.~6c).  As with the
bridge, the FIR emission is displaced slightly from the HI emission.

\subsubsection{IC~1070}

The smaller companion galaxy, IC~1070, has been detected in the radio
continuum (Fig.~1b), infrared (Fig.~3c and
d), and probably in x-ray as well (Fig.~5) but not in
neutral hydrogen.  These images suggest that this galaxy may also be
taking part in the interaction.

\subsection{Disk Emission and Star Forming Regions}
\label{disk_comparison}

In Fig.~7, contours of the disk emission from the different ISM
components are superimposed on the H$\alpha$ greyscale image (except
for Fig.~7a where H$\alpha$ contours are superimposed on the R Band
image).  The CO, radio continuum, and FIR emission (Fig.~3) all peak
at or near the nucleus of NGC~5775, while the HI occurs in a ring with
a central depression.  The optical and H$\alpha$ images show a
prominent dust lane along the eastern side of the major axis (top in
Fig.~7), indicating that this is the closer side.  The north-west (NW)
major axis is advancing and the south-east (SE) side is receding, so
trailing spiral arms will have an integral sign ($\int$) shape for
which there is some evidence in the H$\alpha$ (Fig.~7a), high
resolution radio continuum (Fig.~7b), and CO J=2-1 (Fig.~7e)
distributions.

The 3 brightest knots of H$\alpha$ emission, which we take to
represent the brightest star forming regions, are labelled A, B, and C
(brighter to dimmer) in Fig.~7a.  The first, Feature A, is
very compact and shows the best correspondence with a radio continuum
peak (Fig.~7b).  It is located at the far tip of the SE
major axis ($\alpha$ = 14$^h$54$^m$1\fs1, $\delta$ =
03\degr31\arcmin26\arcsec).  Feature B is the brightest star forming
region near the nucleus, located 6\arcsec\ south of the nucleus (at RA
= 14$^h$53$^m$57\fs6, $\delta$ = 03\degr32\arcmin34\arcsec) and
slightly offset from the radio continuum peak (at $\alpha$ =
14$^h$53$^m$57$\fs$4, $\delta$ = 03\degr32\arcmin43\arcsec; average of
20 cm and 6 cm 5\arcsec-resolution images), likely due to dust
obscuration.  Feature C is located at the far tip of the NW major axis
(at $\alpha$ = 14$^h$53$^m$54\fs5, $\delta$ =
03\degr33\arcmin49\arcsec) at the end of a string of several HII
regions where the radio continuum emission is also stronger.  Several
other HII regions are also visible throughout the disk.

Generally, there is quite good agreement between the 5\arcsec\
resolution 20 cm radio continuum emission and the H$\alpha$ emission
(Fig.~7b), with peaks occurring at roughly the same place
in both maps (a few offsets are also seen).  The CO J=2-1
distribution also correlates with these components
(Fig.~7e), consistent with the fact that these are each
star formation tracers.  The spectral index map (Fig.~7c)
also shows a strong correlation with regions of star formation in the
sense that the spectral index is significantly lower ($\sim$0.6) in
these regions.  Only the disk HI distribution (Fig.~7d)
shows little correlation with the other bands.

Both H$\alpha$ and radio continuum distributions have lower
intensities about 50\arcsec\ NW of the optical nucleus between B and
C, near $\alpha$ = 14$^h$53$^m$55\fs5, $\delta$ =
03\degr33\arcmin20\arcsec. The CO distribution also falls off strongly
as this depression is approached.  The decline in intensity is most
obvious in H$\alpha$ in which there appears to be a ``gap" in the
emission.  The fact that the radio continuum and molecular components
correlate suggests that this H$\alpha$ gap is {\it not} due to dust
obscuration.  Within this region there are also several faint HII
regions, one of which is clearly extraplanar (see
Sect.~\ref{halpha_features}).  At the corresponding offset on the SE
disk between A and B near $\alpha$ = 14$^h$53$^m$59$^s$, $\delta$ =
03\degr32\arcmin00\arcsec, a similar depression in the H$\alpha$
distribution can be seen.  Here, the radio continuum and CO emission
do not fall off as dramatically, suggesting that dust obscuration may
be contributing more strongly to the SE H$\alpha$ decrement.

\subsection{Extraplanar Features}
\label{extraplanar}

NGC~5775 shows numerous extensions, arcs, loops and other connecting
features between the disk and halo in every component of the
interstellar medium.  Even CO, which has not been fully mapped to high
galactic latitudes at the required sensitivity and resolution, shows
some evidence for vertical extensions (Fig.~7e) near the
nucleus.

\subsubsection{High Latitude Neutral Hydrogen}
\label{hi_features}
 
Of the extraplanar emission seen over the various wavebands, it is the
high latitude HI features which are the most well-defined and so we
use these as a reference for comparison with other wavebands.  HI
features in this galaxy have been pointed out by Irwin
\cite*{irwin:94a} and presented in detail in Lee \cite*{lee}.

A total of 6 HI high-latitude features have been identified (Lee
\citeonlyyear{lee}) by visually inspecting the velocity channel
maps. Only features that appear at the same spatial location in at
least two consecutive velocity channels were considered to be real.
In this paper, we concentrate on the three most obvious features,
labelled F1 to F3 in Fig.~1d. All three can be traced
for many channels and reach projected heights of 7.1, 6.2, and 7.3~kpc
from the midplane for F1, F2 and F3, respectively.  F1 and F2 appear
to be roughly symmetrically placed with respect to the galactic
centre. Because of the spatial and velocity coherence of the features,
because they appear loop-like (or arc-like with open tops), and
because they show expansion (see below), we identify them as HI
supershell candidates or remnants, similar to those found in the Milky
Way \cite{heiles:79,heiles:84}.  Their radii, $R_\mathrm{sh}$,
projected galactocentric distances, $D_\mathrm{sh}$, half the velocity
range over which they are observed, $V_\mathrm{sh}$, and number
densities at midplane, $n_\mathrm{o}$, are given in
Table~\ref{tab:5775shellenergy}.  Note that $n_\mathrm{o}$ is
calculated based on the HI modeled density distribution given in Irwin
\cite*{irwin:94a}, assuming that the galactocentric radius is
equivalent to the projected galactocentric radius.  It is interesting
that the most well-defined extra-planar features occur, not near the
nucleus, but at projected radii greater than 4.3~kpc, indicating that
their actual galactocentric radii are greater than or equal to these
values.

Feature F2 shows the clear signature, in position-velocity (PV) space,
of an expanding supershell (see Fig.~9h and Sect.~\ref{F2}).
While the other features show no clear PV expansion signature, we can
infer that expansion is taking place since the velocity range over
which the feature is seen (hundreds of km s$^{-1}$) is greater than
the change in velocity of the underlying rotation curve over the
diameter of the feature at its galactocentric radius (less than 25 km
s$^{-1}$, Irwin \citeonlyyear{irwin:94a}).  In addition, double line
structure is also seen (Sect.~\ref{correlations}).  Note also that there
is no confusion against the disk at these high latitudes.  We can
therefore estimate a kinematic age of the shells, $\tau_\mathrm{sh} =
R_\mathrm{sh}/V_\mathrm{sh}$.  The results (Table~\ref{tab:5775shellenergy}) are of
order 10$^7$ yr, similar in magnitude as those found in NGC~3079
(Irwin \& Seaquist \citeonlyyear{irwin:90}) and NGC~4631 (Rand \& van
der Hulst \citeonlyyear{rand:93}).

If the supershells are produced by supernova explosions in the disk,
we can also estimate the input energies required to form them.  For an
expanding shell which is formed from a one-time energy injection and
which is now in the radiating phase of its evolution, the required energy
injected (Chevalier \citeonlyyear{chevalier}) is $ E_\mathrm{E} =
5.3\times10^{43}n_\mathrm{0}^{1.12}R_\mathrm{sh}^{3.12}V_\mathrm{sh}^{1.4} $
where $n_\mathrm{0}$ is the number density at midplane in cm$^{-3}$,
{\it R}$_\mathrm{sh}$ and {\it V}$_\mathrm{sh}$ are the radius and
expansion velocity of the shell in pc and \kms, respectively. 
Chevalier's numerical models show that $<$ 10\% of the explosion
energy is transferred to kinetic energy of the expanding supershells.
The results, making use of the modeled number density profile given in
Irwin \cite*{irwin:94a} are shown in
Table~\ref{tab:5775shellenergy}. For a typical supernova energy of
10$^{51}$~ergs, the energy input required for F1 alone
(1.9$\times$10$^{55}$~ergs) would require $\approx$ 10$^4$ correlated
supernovae within $\approx$ 10$^7$ yr which is also roughly the age of
a typical young star cluster or OB association.

\begin{table*}
\caption{Parameters for Supershells in NGC~5775}
\label{tab:5775shellenergy}
\begin{center}
\begin{tabular}{llllllll} \hline\hline
\\
Feature & {\it V$_\mathrm{sh}~^a$} & {\it R$_\mathrm{sh}$} & 
{\it D$_\mathrm{sh}$} & {\it n$_\mathrm{0}$} & {\it $\tau_\mathrm{sh,7}~^b$} & 
{\it E$_\mathrm{E,55}~^c$} & {\it E$_\mathrm{c,55}~^d$} \\ 
& \kms & kpc & kpc & \cubiccm & years & ergs & ergs \\ 
(1) & (2) & (3) & (4) & (5) & (6) & (7) & (8) \\
\hline
\\
F1 &  54.0 & 2.2 & 4.3 & 0.07 & 4.0 & 1.9 & 0.7 \\
F2 &  62.5 & 2.0 & 5.0 & 0.08 & 3.2 & 2.1 & 0.9 \\ 
F3 &  75.0 & 1.7 & 7.6 & 0.14 & 2.2 & 3.0 & 2.2 \\
\\ \hline
\multicolumn{8}{l}{$^a$ Half of the total velocity width obtained from
the PV diagrams} \\
\multicolumn{8}{l}{in Figs.~8, 9 and 10. See Sect.~\ref{correlations}.} \\
\multicolumn{8}{l}{$^b$ Kinematical ages in units of 10$^7$~years.} \\
\multicolumn{8}{l}{$^c$ One-time energy injected in units of 10$^{55}$~ergs.} \\
\multicolumn{8}{l}{$^d$ Continuous energy injected in units of 10$^{55}$~ergs.} \\\end{tabular}
\end{center}
\end{table*}

If we consider a slower, continuous energy injection over a typical
cluster ($\sim$shell) age of several $\times$ 10$^7$~years, the energy
required can be expressed as $E_\mathrm{c} =
1.16\times10^{41}R_\mathrm{sh}^5n_\mathrm{0}t_\mathrm{7}^{-3}$~ergs (from Eq.~3 in
McCray \& Kafatos \citeonlyyear{mccray}).  These results are also
listed in Table~\ref{tab:5775shellenergy}. Therefore, the energy
needed to form each supershell, be it a one-time energy injection, or
a continuous supply, is extraordinarily large. NGC~5775 has a high
global star formation rate, i.e. SFR (M $>$ 5 M$_\odot$) = 2.4
M$_\odot$ yr$^{-1}$ (cf. Condon \citeonlyyear{condon}), but creating
the features observed requires that 10$^4$ to 10$^5$ supernovae occur
within the same region on short timescales.  These numbers require the
presence of 10 to 100 coeval and closely located
super star clusters for each supershell, assuming that each
contains 1000 supernova-producing stars.  We return to this issue in
Sect.~\ref{discussion}.

\subsubsection{High Latitude H$\alpha$ Emission}
\label{halpha_features}
 
Since dust obscuration should be minor at high latitude,
 the observed high latitude H$\alpha$ distribution
should closely resemble the actual distribution of ionized gas.
The largest H$\alpha$ disk-halo feature occurs on the east side of the
NW disk up to $\approx$ 5.5 kpc, in projection, from mid-plane
(Fig.~4a) at the location of F3. 
Since the correction for inclination is negligible, this is
 the actual height above the disk, and is much higher than
 the H$\alpha$ ``worms'' found
in NGC~891 \cite{rand:90} and NGC~4631 \cite{rand:92}, which reach to
about 2~kpc.  An unresolved HII region mentioned in
Sect.~\ref{disk_comparison} at high {\it z} ($\sim$1.8~kpc in projection)
occurs at $\alpha$ = 14$^h$53$^m$56\fs5 $\delta$ =
3\degr33\arcmin32\arcsec, just above the H$\alpha$ gap.  Since the
H$\alpha$ disk is about 30~kpc in diameter and {\it i} = 86\degr, a
source at the far edge of the disk will appear at {\it z} = 1.0~kpc in
projection. Therefore, this source is likely extraplanar and appears
to ``feed into" the diffuse, large-scale high latitude feature.  On
the SW disk, there are also two other prominent H$\alpha$ features at
the locations of F1 and F2 which are across the major axis from each
other (Fig.~4a).

\subsubsection{ High Latitude Radio Continuum Emission}
\label{cosmic_rays}

Some of the most dramatic features occur in the radio continuum
(Fig.~1b) in which extensions can be traced as far as
16 kpc, in projection, from the plane.  Irwin \& Caron
\cite*{irwin:94b} and Duric et al. \cite*{duric:98} have shown that
the extraplanar radio continuum emission is largely non-thermal.  The
observed extensions show considerable structure far from the plane and
may be associated with more distant knots of emission which are
disconnected from the disk.

Figs.~1b, 2b and 2d show
contours of the 20 cm continuum data sets at different resolutions and
display the growth of features over spatial scales from 5\arcsec\ =
0.6 kpc to many kpc.  Even with these sensitive data, it is not
straightforward to associate high resolution features with those
observed on larger scales.  Cosmic ray diffusion, possibly complex
magnetic field geometry, and the fact that several features could be
present over any given line of sight in this edge-on galaxy could all
be complicating the interpretation of relationships between small and
large-scale structure.  Nevertheless, some trends can be observed.
Generally, if a feature is seen at high resolution, it 
will also exist farther from the plane
at (at least) the next lower resolution.

For example, the several discrete extensions seen on the west side of
the NW major axis (Fig.~2d) are observed as smooth
extensions at both lower resolutions (same Fig.).  Smaller extensions
on the west side of the SE major axis are also visible at the lower
but not lowest resolutions.  Some of this emission at high resolution
even appears disconnected from the disk (e.g. $\alpha$ =
14$^h$53$^m$55$^s$, $\delta$ = 03\degr31\farcm5) similar to the much
larger scale disconnected features seen at low resolution
(Fig.~1b).

\subsubsection{High Latitude IR Emission}

The FIR emission (Fig.~3) shows several features which
appear to extend from the disk, most easily seen in the 
12~$\mu$m and 25~$\mu$m maps due to their higher spatial resolution.
While it is not always clear whether an extension is due to noise
or a superimposed background source,
there are two which have clear counterparts at other wavebands.
The first is a small feature extending on the SW side of the 
disk at $\alpha$ = 14$^h$53$^m$53\fs5 $\delta$ = 03\degr31\arcmin58\arcsec.  
This feature has a direct HI counterpart (seen in
Fig.~6b) and also correlates with extended
radio continuum emission (Fig.~2a).  The second is 
on the NW side of the disk at
$\alpha$ = 14$^h$53$^m$58\fs2 $\delta$ = 03\degr34\arcmin31\arcsec.
This feature correlates
with HI feature F3 and its associated radio continuum and
H$\alpha$ emission.
The IR emission
occurs at the 3 - 5 $\sigma$ level in each case, but the 
existence of the features at both 12~$\mu$m and 25~$\mu$m 
and the correlation
with the emission at other wavebands argues for this emission being
real.
Weak IR emission at 25 and 60 $\mu$m can
also be seen at the position of one intergalactic radio continuum knot
(see $\alpha$ = 14$^h$53$^m$41\fs8, $\delta$ =
03\degr31\arcmin2\arcsec\ and Fig.~1b). 

While some
correlation has been found in the Milky Way Galaxy between IR
features, radio continuum features and HII regions \cite{koo:92} and
also an HI supershell \cite{kim}, such correlations have not 
previously been readily
apparent in external galaxies.
The FIR features found here are reminiscent of high-latitude FIR in
NGC~891 \cite{alton:98}, M~82, NGC~4631 and NGC~253
\cite{alton:99,garcia}. In these cases, discrete dusty features are
detected out to about 1~kpc in {\it z}. Extraplanar dust detected
optically is also found to be common in a sample of 12 edge-on spiral
galaxies by Howk \& Savage \cite*{howk}. In the Milky Way Galaxy,
smaller examples of ``dusty worms'' are found in Koo et
al. \cite*{koo:91}. The detection of FIR emission at high-{\it z}
suggests that the dust is being transported upwards away from the
plane with the other components without being destroyed. Various
transport mechanisms including radiation pressure and hydrodynamical
effects are discussed in Howk \& Savage \cite*{howk}.  

\subsubsection{High Latitude X-ray Emission}

The x-ray image (Fig.~5) shows 3 main regions in which the
emission extends in discrete features, two on the east side of the
galaxy and one on the west.  As the observations are on a much larger
spatial scale than that over which some of the disk-halo features are
observed, it is not straightforward to relate the x-ray features
directly to other known disk-halo features.  However, it is of
interest to note that the three x-ray extensions occur roughly at the
positions of disk-halo features, F1, F2, and F3.  In the case of F3,
since this feature can be traced to very large scales in the radio
continuum (cf. Fig.~1b), it is possible to associate
this disk-halo feature (with more confidence) with the northern x-ray
feature on the east side of the galaxy.  Since the x-ray resolution is
30\arcsec, the highest latitude x-ray emission, if real, is associated
with F3 and occurs at $\sim$ 17 kpc, in projection, from the plane,
similar to the radio continuum.

\subsection{Relationship Between In-Disk and
Extraplanar Features}
\label{disk_highz_comparison}

Fig.~7 shows in-disk features (especially H$\alpha$) in
comparison with other wavebands.  As there are many extraplanar
features, the relationship between those in the disk and those above
in this edge-on system are not easily established.  However, F3, which
is seen in {\it every} waveband (except for CO for which we have
insufficient coverage), is clearly located above a region of {\it low}
intensity on the major axis between features B and C (as labelled in
Fig.~7a).  In this region, all sufficiently resolved
components of the ISM, except for HI, are at a minimum. There is also
evidence for extensions away from this feature on the opposite side of
the major axis though we have avoided this side due to its proximity
to the southern bridge.  Thus, for F3, there is an anti-correlation
between the high latitude emission and tracers of in-disk star
formation, especially in the radio continuum where the intensity dip
is not an effect of dust absorption.  Extraplanar features F1 and F2
also occur on opposite sides of the SE major axis between A and B
where the in-disk H$\alpha$ is weaker but the radio continuum is still
fairly strong (Sect.~\ref{disk_comparison}).

\subsection{Correlation of Extra-Planar Features at Different
Wavebands}
\label{correlations}

Figs.~8, 9, and 10 show overlays of all
wavebands with sufficient and similar resolutions for the three HI
features, F1, F2, and F3.  The HI features may appear slightly
different from Fig.~1c and d, since they are here
averaged only over the velocity range over which they occur.

\subsubsection{F1}
\label{F1}

Fig.~8 shows a complete HI loop of radius 2.2 kpc.  The
velocity field (Fig.~8b) shows that the bulk of the gas is
blueshifted with respect to gas on the major axis.  Since this is the
receding side of the disk, F1 is lagging behind the general rotation
of the galaxy disk.  This velocity offset is confirmed from
position-velocity slices shown in panels (g), (h), and (i). These
three panels show slices perpendicular to the major axis,
corresponding to the left, center, and right positions shown in (a) (V
= 0 represents the systemic velocity).  The maximum displacement for
``connected" emission between the high latitude gas and gas on the
major axis is $\sim$ 108 km s$^{-1}$, but the velocity offset would be
less if F1 is located at some position other than the projected
galactocentric distance. While F1 shows a complete loop structure in 
position-position space, there is no clear loop structure in
position-velocity space, as would be expected if this were an expanding
supershell.  Rather, there is
 an interesting double velocity
structure evident in the PV slice of Fig.~8i and hinted at
in panels (g) and (h).  A spur can be seen at 200 km s$^{-1}$ which
suggests that redshifted (moving faster than the disk) gas may also be
present.  This spur appears to extend into some low intensity peaks
which move off diagonally towards higher {\it z} and higher positive
velocity.  If this feature is real, then it suggests the presence of
gas which is {\it accelerating} with {\it z} height, with the velocity
offset becoming greater than 300 km s$^{-1}$ by {\it z} = 50\arcsec\
(6 kpc).  The suggested velocity gradient is 6.75 km s$^{-1}$
arcsec$^{-1}$ or 56 km s$^{-1}$ kpc$^{-1}$ up to a height of {\it z}
$\sim$ 6 kpc.  Other faint features can be seen over a wide velocity
range at heights up to 140\arcsec\ (17 kpc) and may form an arc in PV
space (center slice).  These features exist at the 3$\sigma$ level.

Fig.~8c shows the H$\alpha$ emission in greyscale compared
with HI contours.  The H$\alpha$ emission reaches high {\it z} here
and is primarily confined {\it within} the HI loop. (Note that
the {\it apparent} second feature to the NW (right) is due to a
stellar artifact.  There is only one discrete H$\alpha$ extension in
this figure.)  While primarily within the HI loop, some
H$\alpha$ emission extends to higher {\it z} than the HI loop
suggesting, along with the PV slices, that the HI is not
closed at the top even though it appears closed in panel (a).

Both the 6~cm and 20~cm radio-continuum emission at the position of F1
are extended as seen in Fig.~8d and e, but the emission
appears smoother than the HI (even though the resolution is similar)
because of cosmic ray diffusion.  

The spectral index map shows more structure than either of the radio
continuum maps and may prove to be a more useful tracer of activity
than the intensity images.  For example, the spectral index is steeper
{\it within} F1 and less steep along the eastern and western limbs of
the supershell (Fig.~8f) (see also Collins et
al. \citeonlyyear{collins}).  Flatter spectral indices can result
either from intrinsic flattening of the non-thermal index, such as
would occur in shocked regions, or because of a contribution from
thermal gas.  Since we have the H$\alpha$ images, it is possible to
distinguish between these scenarios.  Fig.~8c, for
example, shows most H$\alpha$ emission to be within the loop.  If
the spectral index changes are due to a thermal contribution, then
we would expect a flattening within the shell, rather than along the
rims, as is observed.

This can be further quantified
 by estimating how much emission would be required
from thermal gas alone to flatten the spectral index by the
amount observed ($\sim$ 0.2).  We take the non-thermal spectral index,
$\alpha_\mathrm{NT}$, to be the value within the supershell where the
spectral index is steep (i.e. 1.13).  Then the required thermal
contribution to the flux, $S_\mathrm{T}(\nu_1)$, at frequency, $\nu_1$ 
($S_\mathrm{\nu}\,\propto\,\nu^{-\alpha}$) in the ridge is given by
$$
S_\mathrm{T}(\nu_1)\,=\, {
{S_\mathrm{Tot}(\nu_1)\,-\,{\big({{\nu_1}\over{\nu_2}}\big)}^{-\alpha_\mathrm{NT}}\,S_\mathrm{Tot}(\nu_2)}
\over
{1\,-\,{\big({{\nu_1}\over{\nu_2}}\big)}^{-\alpha_\mathrm{NT}+0.1}}
}
$$
where $S_\mathrm{Tot}(\nu_i)$ is the measured flux in the ridge at frequency,
$\nu_i$.  For the western ridge, we find that a thermal contribution
of $S_\mathrm{T}(1.45~GHz)$ = 0.083 mJy beam$^{-1}$ is required.  The electron
density, $n_\mathrm{e}$, can then be calculated (cf. Mezger \& Henderson
\citeonlyyear{mezger}) for an adopted geometry.  In the simple case of
cylindrical geometry, constant density, and an electron temperature of
10$^4$ K, this is
$$
{{n_\mathrm{e}}} \,=\, 543\,
{{\nu}}^{0.05}\, 
{{{S_\mathrm{T}(\nu)}}}^{0.5}\, 
{{\cal D}}^{-0.5}\,
{{\theta}}^{-1.5}\,~~~~~cm^{-3}
$$
where $\cal D$ is the distance to the source (in kpc), $\theta$ is the
source diameter (arcminutes) and $\nu$, ${S_\mathrm{T}(\nu)}$ are in GHz and
Jy, respectively. The constant has a weak dependence on frequency,
electron temperature and geometry, but varies by no more than 25\%
over a wide range of values.  Taking $\theta$ = $\theta_\mathrm{beam}$ =
0\farcm24, and the above thermal flux, we find $n_\mathrm{e}$ = 0.28 cm$^{-3}$.
The emission measure, $EM$, over a path length, $l$, equal to the beam
size, is then $EM$ = $\int {n_\mathrm{e}}^2 \,dl$ $\approx$ 132 cm$^{-6}$ pc.
The H$\alpha$ intensity at 10$^4$ K would be $\int I_\nu\, d\nu$ and
is equivalent to 2.8 $\times$ 10$^{-16}$ erg s$^{-1}$ cm$^{-2}$
arcsec$^{-2}$.  This is an order of magnitude greater than the actual
H$\alpha$ value measured at this position of 1.8 $\times$ 10$^{-17}$
erg s$^{-1}$ cm$^{-2}$ arcsec$^{-2}$ and thus a greater thermal
contribution than is observed is required to flatten the spectral
index.

We conclude that the spectral index flattening is {\it not} due to a
thermal contribution, but rather is due to real differences in the
non-thermal spectral index.  This conclusion does not change if we
make reasonable alterations to the geometry or positions chosen for
measurement.  Given that the radio continuum brightness in the ridges
is also greater, the spectral index has probably been flattened in
shocked regions.  Rand \cite*{rand:00} has suggested that the optical
line ratios in high latitude gas cannot be explained by
photoionization alone, but require either a changing temperature or a
contribution from shocks.  These results are consistent with a shock
explanation.

\subsubsection{F2}
\label{F2}

The second feature, F2, is on the SE major axis, extending towards the
west (Fig.~1d) and is shown in Fig.~9. As
with F1, this feature has a loop-like appearance with an additional
feature extending to the west.  The velocity field of this supershell
(Fig.~9b) is also slightly blueshifted with respect to the
disk below.  Since this part of the disk is receding from the
observer, the bulk of the emission from F2 is lagging with respect to
disk rotation.  Slices perpendicular to the major axis, shown in
panels (g), (h), and (i), confirm this result but also reveal a more
complex structure.  The center slice clearly reveals the signature of
an expanding shell in PV space with a full width of 125 km s$^{-1}$.
This feature appears to have a closed top.  The left and right slices,
however, both show a double structure, similar to that seen in F1 but
more obviously.  The left slice shows a redshifted spur in the
direction of a disconnected feature which again, suggests acceleration
with {\it z} height.  In this case, the gradient is 5.0 km s$^{-1}$
arcsec$^{-1}$ or 41 km s$^{-1}$ kpc$^{-1}$ to a height of 45\arcsec\
or 5.4 kpc.  The right slice shows a clear split profile with a
velocity separation of $\sim$ 165 km s$^{-1}$ and no evidence for
acceleration.  These features have open tops.  It is possible that we
are observing the superposition of 3 different features along this
line of sight.  However, the loop-like appearance of the feature in
position-position space, the loop-like spectral index distribution
(see below), the location of F2 across the major axis with respect to
F1, and the absence of similar striking features elsewhere (aside from
F3) suggests that we are mainly observing a single feature.

As in F1, the H$\alpha$ emission (Fig.~9c) reaches high
latitude and occurs mostly within the supershell. At the centre of the
supershell where it is connected to the disk, an obvious H$\alpha$
``worm'' can be seen amidst other smaller discrete vertical features.

The 6~cm contours in Fig.~9d show an inverted $T$-shaped
plume extending from one side of F2 and almost connects with an
extension west of it.  This feature, and some others, bears an
interesting resemblance to those seen on the sun.  In contrast, the
20~cm contours show a single discrete extension centered on F2 and
another one which corresponds to the 6~cm extension in the western
edge of the frame.  The spectral indices (Fig.~9f) steepen
within F2 coincident with the H$\alpha$ worm.  As in F1, the eastern
and western limbs of F2 as well as the top show a striking flattening
of the spectral index.  A similar analysis to F1 indicates that,
again, this flattening must be intrinsic to the non-thermal spectrum,
suggesting the presence of shocks.

\subsubsection{F3}
\label{F3}

The disturbance related to F3 (Fig.~10) is the largest and
most obvious in this galaxy. The HI map shows the supershell to be
open-topped, unlike the projected appearance of F1 and (to some
extent) F2.  The velocity field shown in Fig.~10b is chaotic
but is generally redshifted with respect to the disk gas.  Since the
underlying disk is advancing, the bulk of the gas lags behind the
disk.  This is again illustrated in the PV slices shown in panels
(g), (h), and (i).  There is only a hint, in panel (i), of double
structure showing some blueshifted gas. If the blueshifted feature is
real, then the velocity separation is $\sim$ 150 km s$^{-1}$.  All of
the slices show open tops and also show many smaller features over a
wide velocity range at {\it z} $\sim$ 60\arcsec\ (7.2 kpc).  It is of
interest to examine the left slice, especially, since this is roughly
where, in optical emission, Rand \cite*{rand:00} has measured a steady
decelerating velocity gradient of $\sim$ 22~km s$^{-1}$ kpc$^{-1}$
with {\it z} until {\it z} $\sim$ 5 kpc after which the velocity
remains roughly constant at 1660 km s$^{-1}$, or -20 km s$^{-1}$ with
respect to systemic (Slit 1 in Rand \citeonlyyear{rand:00}).  From
Fig.~10g, we find similar behaviour.  A curve through the
brightest emission shows decreasing velocities until {\it z} $\sim$
40\arcsec\ (5 kpc) and then remains constant near V$_\mathrm{sys}$.  The
velocity gradient is $\sim$ 17~km s$^{-1}$ kpc$^{-1}$ which we take to
be consistent with Rand's value, given the fact that this gradient
shows some curvature, and the fact that the resolutions of the two
data sets are quite different.  Note also that Fig.~8g,
through F1, shows a similar trend as Rand's Slit 2 (HI velocity
gradient of about 17~km s$^{-1}$ kpc$^{-1}$ versus Rand's 21~km
s$^{-1}$ kpc$^{-1}$), although Slit 2 is offset by about 22\arcsec\
from F1.

In Fig.~10c, the H$\alpha$ emission again fills the inside
edge of the supershell but, in this case, the H$\alpha$ emission
clearly shows its own double open-topped structure and occurs on the
{\it inner rim} of the HI features. This, along with the similarity in
H$\alpha$/HI velocity structure along the left slice suggests that the
H$\alpha$ emission represents the inner ionized skin of the HI
supershells.

Correlations between the 6~cm, 20~cm emission and HI are difficult to
discern because the radio continuum emission displays many extensions
in this region. It is noteworthy that the radio continuum emission
shows disconnected pieces at high latitude, reminiscent of the
inverted T-shaped piece in F2 after being detached from the
galaxy. For F3, the spectral index does not show as clear a signature
as the other two features, although a ridge of flatter spectrum
emission can be seen in a feature to the left.

\subsubsection{Higher Latitude Emission}

It is very interesting that, for F1, F2, and F3, there appears to be a
change of character in the disk-halo features at {\it z} heights of
between $\sim$ 50 and 60\arcsec\ or $\sim$ 6 to 7 kpc.  At this
height, the features appear to break up into many smaller features
which, individually, are at low S/N, but form coherent enough
structures in PV space that they appear to be real.  For example,
related to F1, a curved string of emission features is seen at heights
up to {\it z} $\sim$ 140\arcsec\ or 17 kpc (Fig.~8h) which
is roughly the maximum height observed in any component related to
this galaxy.  F3 (Fig.~10h) also shows a string of HI
features reaching {\it z} $\sim$ 65\arcsec, or 8 kpc.  Similar
PV features have also been observed in the galaxy, NGC~2613, above
high latitude HI extensions (Chaves \& Irwin \citeonlyyear{chaves}).

These features occur over a very wide range of velocities, both
positive and negative with respect to galaxy rotation.  This is
consistent with the implication from the HI velocities that both
redshifted and blueshifted gas may be present, even though the lagging
gas dominates at low latitudes.  Although we have avoided the 
region on the other side
of the major axis from feature F3 due to its proximity to the southern
bridge, it is worth showing, in Fig.~11 a single PV
slice at the same position as Fig.~10h but extending to the
south of the major axis as well.  The individual features occur on
both sides of the major axis and at almost all velocities.  The
impression is that, around {\it z} of 5 - 7 kpc, the feature has
become unstable and broken up into many cloudlets which exist over
a wide range of velocities more typical of Population II objects.

\subsection{Anatomy of an HI Supershell}
\label{anatomy}

Based on the above results, we have put together a picture of a
`typical' HI supershell in NGC~5775.  This is shown in
Fig.~12.  The particular display is most relevant to
feature, F2, but the general appearance will be similar for the
others, though the presence and extent of the cap and sides will vary.
The inner rims of the HI walls will be ionized and shocks will also
exist within the walls, flattening the spectral index there.  Hot,
x-ray emitting gas will also be present.  In this picture, the feature
is formed by an outflowing conical wind and the HI accelerates
in both the redshifted and blueshifted direction with respect to
underlying rotation.  At higher {\it z}, the lagging side of the feature
dominates and a coherent structure is less obvious as the supershell
breaks up.  At the highest {\it z}, the feature may be partially or
fully open and smaller clouds of HI continue to be present.

\section{Discussion}
\label{discussion}

Our multi-wavelength approach has proven to be fruitful and, in
fact, necessary in order to understand the activity in this galaxy.
Information on each component of the ISM has allowed us to piece
together a view of the disk-halo interaction with unprecedented detail.  
Some important observations are that a) all
components of the ISM 
(with the exception of CO for which there is insufficient
spatial coverage) are represented in a disk-halo feature, 
b) there is a minimum in the emission suggesting a lower level
of star formation
in the disk under the largest feature, F3,
c) the 
bulk of the HI in discrete extensions tends to
 lag with respect to the underlying
disk and d) the 
velocities are suggestive that gas is accelerating with higher {\it z}
and clearly show the presence of an expanding supershell in the case of
F2. 
These observations suggest that we are dealing with {\it
outflows} and rule out inflows for the observed features. 
We seem to be viewing the remnants of supershells as they expand
and break up through the disk-halo interface.

Based on the features seen in NGC~5775, we propose a generic 
disk-halo structure such as is 
 illustrated in Fig.~12.  In this
picture, gas accelerates as it ascends into the halo, possibly from
cone-like outflow.  The split line profiles or expansion signatures in
PV space indicate a classical supershell structure.  However, the
features are either open-topped or only partially capped and ionized
gas, as seen in H$\alpha$ emission, escapes out of the top.  The
ionized gas also forms an inner skin on the HI as shown spatially as
well as in velocity structure which is consistent with the HI. 
The existence and extent of the cap may vary.  Thus, the signature
of a complete ``supershell" may or may not be present. 
 Both red and blueshifted gas are present in these
features as well as evidence for acceleration away from the
plane. However, it is the lagging gas which dominates.  The
possibility of a 
globally lagging halo will be investigated in a separate
paper. At higher {\it z} there appears to be a transition region
between 5 and 7~kpc in which the shell breaks up and HI clumps are
seen over a wide range of velocities.  

Let us now review the possible formation mechanisms and consider
which, if any, may be applicable to NGC~5775.

\subsection{Supernovae}
\label{supernovae}

The effect of supernova explosions on the ISM has been widely studied
both observationally in our own and other galaxies
(e.g., Katgert \citeonlyyear{katgert},
Heiles \citeonlyyear{heiles:79},
Westerlund \& Mathewson \citeonlyyear{westerlund},
Brinks \& Bajaja \citeonlyyear{brinks})
and theoretically 
(Bruhweiler et al. \citeonlyyear{bruhweiler}, Tomisaka et al
\citeonlyyear{tomisaka}, Mac Low \& McCray \citeonlyyear{maclow}).
Galactic
``fountain" and ``chimney" models (Shapiro \& Field
\citeonlyyear{shapiro}, Norman \& Ikeuchi \citeonlyyear{norman})
suggest that supershells are produced from multiple supernovae, both
temporally and spatially correlated.  These shells may then break out
of the disk and provide channels through which hot gas can escape into
the halo.

In the case of NGC~5775, the high mass star formation rate (SFR)
inferred by the FIR luminosity (2.6$\times$10$^{10}~L_{\sun}$, using
the equation given in Lonsdale Persson \& Helou
\citeonlyyear{lonsdale}) is 2.4~$M_{\sun}\mathrm{yr}^{-1}$ (from
equation 26 in Condon \citeonlyyear{condon}). This is comparable to
the SFR in M~82 (Young \& Scoville \citeonlyyear{young}). It is
therefore conceivable that massive stars play an important role in
shaping the ISM in this galaxy. Kennicutt et al. 
\cite*{kennicutt} find that the brightest HII region
in normal late-type galaxies contains the equivalent of about 20,000 O
and B stars, though very few galaxies have such
large associations.
Over a hundred super star
clusters (SSCs) have also been discovered in the starburst nucleus of
M~82 using the Hubble Space Telescope.
 Some of these are extremely
luminous (M$_\mathrm{V}\sim$-16$^m$) and massive ($\sim$10$^7$~$M_{\sun}$) with
ages comparable to the ages of the supershells in NGC~5775
(O'Connell et al. \citeonlyyear{oconnell}, Gallagher \& Smith
\citeonlyyear{gallagher}).  While we do not have information on
the SSC population in NGC~5775, the limitations of supernovae alone
as an explanation for the observed supershells are that a) a very large
number 
(10$^{4}$ to 10$^{5}$) of correlated supernovae are required and b)
there is little evidence for such activity 
at the base
of feature F3.  
If activity related to massive star formation
 has indeed formed F3, then it must have declined rapidly
 within the lifetime of the feature ($\sim\,2.2\,\times\,
10^7$ yrs).
 
Several recent studies have shown that when magnetic fields are
included in numerical simulations not just as a secondary effect, then
large shells and blow-out can occur with more modest energy
requirements (e.g., Kamaya et al. \citeonlyyear{kamaya}). 
In the process, the field lines drag the partially ionized
gas, dust and energetic cosmic rays along to high latitude.
 Sofue et al. \cite*{sofue} find that this ``magnetic flux
inflation'' can be used to explain the multitude of out-of-plane dust
features in NGC~253. 
 Thus, magnetic instabilities could exist as a
non-negligible enhancement to the primary supernova mechanism for
producing HI supershells.

\subsection{Hypernovae}
\label{hypernovae}

Another possible energy source for supershell formation, which
alleviates the requirement of extraordinarily large clusters, is
hypernovae (see also Efremov et al. \citeonlyyear{efremov},
Paczynski \citeonlyyear{paczynski}) which may be able to release 
over 10$^{54}$~ergs of kinetic energy and have been proposed
as a possible progenitor of gamma-ray bursters (GRBs).  
The frequency of GRBs, assuming they are powered by hypernovae, is
10$^4$-10$^5$ times less common than supernovae \cite{paczynski}.  
We estimate the supernova rate from Condon \cite*{condon},
$\nu_\mathrm{SN} = 3.7\times10^{-12}L_\mathrm{FIR}\,=\,0.1~yr^{-1}$,
leading to a hypernova rate of about
10$^{-5}$-10$^{-6}$~yr$^{-1}$. We therefore expect to see 10-100
hypernovae within a period of 10$^{7}$~years
or the lifetime of the HI features.  Since 6 HI supershells
have been identified in Lee \cite*{lee}, this is roughly consistent
with the proposed hypernova rate, though it should be kept in mind 
that the characteristics of such objects, if they exist, are not
well known.

\subsection{External Cloud Impacts}

The impact of high velocity clouds (HVCs) with the disk has been
proposed as the formation mechanism of some HI supershells in our own
Galaxy \cite{mirabel,meyerdierks} and in external galaxies in which
input energies are large.
While we have argued (Sect.~\ref{correlations}) that the features
seen in NGC~5775 represent outflows, it is interesting to note that
outflow features are also predicted in the infalling cloud scenario,
for example, an impact with a magnetized disk can produce outflowing
velocities $>$ 100 km s$^{-1}$ and acceleration with {\it z} height
(Santill\'{a}n et al. \citeonlyyear{santillan}). 

Since NGC~5775 is clearly interacting with at least NGC~5774, a source
of clouds is potentially present.  To help determine whether impacting
clouds can explain the outflow on energetic grounds, we adopt an 
impacting cloud mass of $2\times10^7~M_{\sun}$, similar to that found 
by Rand \& Stone \cite*{rand:96}, and a {\it z} height of 35~kpc as a 
typical initial position (corresponding to the HI emission feature
seen in Fig.~1d at $\alpha$ = 14$^h$53$^m$40$^s$, $\delta$ =
3\degr30\arcmin30\arcsec).  The potential energy of such
a cloud is (Binney \& Tremaine \citeonlyyear{binney})

\begin{eqnarray} 
\Phi(\mathrm{z}) & = & 2\times10^{54}\left(\frac{M_\mathrm{cloud}}
{2\times10^7~M_{\sun}}\right)\left(\frac{z_\mathrm{o}}{700~\mathrm{pc}}\right)^2
\cdot \nonumber \\
& & \left(\frac{\rho_\mathrm{o}}{0.185~M_{\sun}\mathrm{pc}^{-3}}\right)
ln\left[cosh\frac{z}{z_\mathrm{o}}\right] \nonumber,
\end{eqnarray}

\noindent where  $z_\mathrm{o}$ is the scale height of the vertical
distribution of stellar mass and $\rho_\mathrm{o}$ is the stellar mass density
at the midplane. The last two parameters take the Galactic and solar
neighbourhood values, respectively.
 We can
then equate the kinetic energy at the point when the cloud hit the
galactic plane ($\frac{1}{2}M_\mathrm{cloud}v^2$, where $v$ is the velocity
at the point of contact) with the potential energy to obtain the
velocity of impact. This gives $v$ = 66~\kms. Santill\'{a}n et al.
\cite*{santillan} have shown that the resulting outflowing gas
velocities after impact can achieve 64\% to 100\% of the infalling
cloud velocity.  Thus, this value is consistent with the expansion
velocities of the supershells observed here.  Consequently, on
energetic grounds, impacting clouds could be responsible for the
observed features.

However, for NGC~5775, this model has more difficulty accounting for
the velocity structure of the features and their symmetry.  For example,
in the models of Santill\'{a}n et al.,
outflows are produced when clouds ``rebound" from the disk via
magnetic tension.  Thus, the cloud is not expected to penetrate the
disk and outflows would not be seen on both sides of
the disk (as with F1, F2) unless there happens to be an infalling cloud 
symmetrically on the other side of the disk as well.  In any case, one
would not expect to see closed loops in projection (e.g. F1) from such
``splashback" and one would not expect to see the signature of an
expanding shell in PV space (e.g. F2).  If an impacting cloud somehow
passed through the disk, again, it is unlikely that there would be
evidence for outflow on both sides.
 Finally, high velocity
cloud impacts with a velocity less than 300~\kms\ cannot shock-heat gas
into x-ray-emitting temperature. If the x-ray emission from the
high-{\it z} features is real, we can exclude this mechanism. Thus,
even though NGC~5775 is an interacting galaxy, the anatomy of the
disk-halo features in NGC~5775 is {\it inconsistent} with the
impacting cloud scenario.

\section{Summary and Conclusions}

In this paper, we examined the distribution of the various ISM
components in the edge-on, infrared-bright, interacting galaxy,
NGC~5775 which shows an active disk-halo interface. All components of
the ISM are presented, including emission from HI, the radio
continuum, CO, FIR, X-ray and H$\alpha$.
 NGC~5775 contains at least 3 well-defined HI disk-halo
features which we have interpreted as representing supershells or
remnants of supershells as they evolve through the disk-halo interface. 

NGC~5775 is also interacting with its companion galaxy, NGC~5774, 
(and possibly IC~1070 as well) and
this is manifested in the form of two bridges connecting the two
galaxies, as well as a possible tidal tail at the southeastern tip of
NGC~5775. In particular, the southern intergalactic bridge can be
traced in all observed wavebands except CO for which data are not yet
available.

At least three obvious extraplanar features (F1, F2 and F3) can be
traced in HI. The largest of these (F3) reaches a projected height of
7.3~kpc.  The energy required to produce each of these features is
high, of order 10$^{55}$~ergs, or equivalent to 10$^4$
supernovae. High latitude emission associated with such features can
be seen in {\it all} wavebands, representing all components of the ISM
(except CO for which there is insufficient coverage).  This includes
the FIR emission, indicating the presence of dust at high latitudes as
well. There is a {\it minimum} in the disk emission beneath the most
obvious disk-halo feature, F3.  Evidence for acceleration away from
the plane and double line velocity structure with both red and
blueshifted sides are seen. One feature (F2) shows the clear signature
of an expanding supershell in PV space.  The others do not, but are
suggestive of parts of a supershell which have broken up.

H$\alpha$
emission is located interior to the HI and exhibits similar velocity
structure where measured (F3). From the radio continuum data, we have
clearly shown that the flattening of the spectral indices at the
ridges of the features are due to shocks and cannot be
attributed to a contribution from thermal gas. 

Based on the above results, a picture of a typical HI supershell is
presented where the supershell is created by an outflowing conical
wind. X-ray emitting gas
and ionized gas exist within the HI supershell. At high-{\it z}, the
parts of the features that are lagging with respect to the galactic
rotation dominate (see Fig.~12). At higher {\it z}, 
numerous smaller clumps at both red and blueshifted velocities
are seen.

Although NGC~5775 is interacting, the symmetry 
 and velocity structure of the observed
disk-halo features are inconsistent with the impacting cloud
scenario. These features are likely internally generated via a violent
process, given the energies and shocks observed.  If supernovae are
the origin of the features, then of order 10$^4$ correlated supernovae
are required and must fade within a time period which is shorter
than the lifetime of the features.

\medskip

{\bf Note Added in Proof:}
{Since the original
manuscript of this paper was written, we have made clear detections
of CO(1-0) and CO(2-1) in F2 and also 
850$\mu m$ dust emission in F1, F2, and F3.}

\begin{acknowledgements}
SWL would like to thank Dr. D. Wing for the many useful discussions
related to this paper. JI wishes to thank the Natural Sciences and
Engineering Research Council of Canada for a research grant. This work
has made use of the Digitized Sky Survey of the Space Telescope
Science Institute.  The Space Telescope Science Institute is operated
by the Association of Universities for Research in Astronomy, Inc. for
the National Aeronautics and Space Administration.  The Digitized Sky
Survey was produced under Government grant NAG W-2166.  We have also
made use of the NASA/IPAC Extragalactic Database (NED) which is
operated by the Jet Propulsion Laboratory, under contract with the
National Aeronautics and Space Administration.
\end{acknowledgements}

%\begin{thebibliography}{}
\bibliography{aamnem99,ngc5775}
\bibliographystyle{aabib99}

\newpage

\begin{center}
{\large Figure Captions}
\end{center}

Fig. 1: Sequence of images of NGC~5775 (edge-on), NGC~5774 (nearly
face-on, to the north-west), and IC~1070 (to the south).  The optical
positions of each galaxy, from the NASA/IPAC Extragalactic Data Base
(NED), are marked with white or black crosses.  The synthesized beam
(if applicable) is shown at the bottom left of each image.  {\bf a)}
Optical image from the Space Telescope Science Institute (STScI)
Digitized Sky Survey (DSS).  {\bf b)} VLA 20 cm radio continuum image
from combined B, C, and D array data (see Duric et al. 1998).  The
synthesized beam is 23\farcs5 $\times$ 23\farcs4 at a position angle
(PA) of 45\fdg2.  The rms noise is 43 $\mu$Jy beam$^{-1}$, the peak is
34.9 mJy beam$^{-1}$ and contour levels are at $\pm$0.065, $\pm$0.1,
$\pm$0.2, 0.4, 0.75, 2.0, 5.0, 10, 20, and 34 mJy beam$^{-1}$.  {\bf
c)} Column density map of the naturally weighted neutral hydrogen
emission (see Irwin 1994). The beam is 28\farcs8 $\times$ 22\farcs8 at
PA = 52\fdg0, the peak is 72 $\times$ 10$^{20}$ cm$^{-2}$, and
contours are at 0.5, 1, 2, 3, 5, 10, 15, 20, 30, 40, 50, 60, 70, and
80 $\times$ 10$^{20}$cm$^{-2}$.  {\bf d)} As in {\it c)} but using
uniformly weighted data. The beam is 13\farcs6 $\times$ 13\farcs4 at
PA = -33\fdg7, the peak is 100 $\times$ 10$^{20}$ cm$^{-2}$ and
contours are at 1, 5, 10, 17.5, 25, 40, 60, and 82.5 $\times$
10$^{20}$ cm$^{-2}$.  HI extensions are labelled F1 to F3.\\

Fig. 2: {\bf a)} VLA 6~cm radio continuum image of NGC~5775 from
combined C and D array data (see Duric et al. 1998). The restoring
beam is 15\farcs26 $\times$ 13\farcs50 at PA = 20\fdg07. The map peak
is 8.7~mJy beam$^{-1}$ and rms noise is 20~$\mu$Jy beam$^{-1}$.
Contour levels are at $\pm$0.03, $\pm$0.06, 0.09, 0.15, 0.3, 0.75,
1.5, 3.0, 4.5, 6.0, and 7.5~mJy beam$^{-1}$.  {\bf b)} VLA 20~cm radio
continuum image of NGC~5775 from combined B, C and D array data (see
Duric et al. 1998). The restoring beam is 15\farcs26 $\times$
13\farcs50 at PA = 20\fdg07. The map peak is 19.1~mJy beam$^{-1}$ and
rms noise is 25~$\mu$Jybeam$^{-1}$.  Contour levels are at $\pm$0.07,
0.14, 0.21, 0.35, 0.7, 1.05, 1.75, 3.5, 7.0, 10.5, and 14~mJy
beam$^{-1}$.  {\bf c)} The spectral index map made by combining the
maps from 2a and 2b (see Duric et al. 1998). The spectral index,
$\alpha$, is defined according to the S($\nu$)$\propto \nu^{-\alpha}$
convention. The contours start at 0.6 (light grey) to 1.0 (black) with
an increment of 0.05.  {\bf d)} VLA 20~cm radio continuum image of
NGC~5775 from the B array data (see Duric et al. 1998). The restoring
beam is 5\farcs03 $\times$ 4\farcs96 at PA = -41\fdg48. The map peak
is 4.2~mJy beam$^{-1}$ and rms noise is 30~$\mu$Jy
beam$^{-1}$. Contour levels are at $\pm$0.09, 0.15, 0.3, 0.6, 1.5,
2.1, and 3.0~mJy beam$^{-1}$.  \\

Fig. 3: As in Fig. 1 but showing the HIRES images.  {\bf a)} 12 $\mu$m
image. The beam is 59\farcs4 $\times$ 30\farcs6 at PA = 15\fdg6.  The
map peak is 11.5 MJy sr$^{-1}$ and the rms noise is 0.29 MJy
sr$^{-1}$.  Contours are at $\pm$0.44, $\pm$0.58, 1.0, 1.8, 3.0, 4.0,
and 7.0 MJy sr$^{-1}$.  {\bf b)} 25 $\mu$m image.  The beam is
57\farcs6 $\times$ 30\farcs6 at PA = 15\fdg7.  The peak is 14.1 MJy
sr$^{-1}$ and rms noise is 0.38 MJy sr$^{-1}$.  Contours are at
$\pm$0.57, 0.8, 1.2, 2.3, 3.5, 5.7 and 9.0 MJy sr$^{-1}$.  {\bf c)} 60
$\mu$m image.  The beam is 75\farcs0 $\times$ 46\farcs8 at PA =
16\fdg8.  The peak is 128.8 MJy sr$^{-1}$ and rms noise is 0.14 MJy
sr$^{-1}$.  Contours are at $\pm$0.24, $\pm$0.33, 0.65, 1.3, 2.6, 6.4,
13, 26 and 58 MJy sr$^{-1}$.  {\bf d)} 100 $\mu$m image.  The beam is
101\farcs4 $\times$ 82\farcs2 at PA = 19\fdg8.  The peak is 200.4 MJy
sr$^{-1}$ and rms noise is 0.15 MJy sr$^{-1}$.  Contours are at
$\pm$0.23, $\pm$0.50, 1.0, 2.0, 4.0, 10, 20, 40, 80 and 125 MJy
sr$^{-1}$.\\

Fig. 4: {\bf a)} The continuum-subtracted H$\alpha$ image of
NGC~5775. Seeing is about 1\arcsec.  Contours are 4, 5, 10, 20, 50,
100, 150, 250, 500 $\times$
6.6$\times$10$^{-18}$~\ergpers~cm$^{-2}~$arcsec$^{-2}$.  The white
cross marks the optical position of the galaxy.  {\bf b)}
$^{12}$CO~J=1-0 integrated intensity map of NGC~5775 obtained at the
SEST. The beam size is 43\arcsec\ and the map peak is
22.6~K~kms$^{-1}$. Contour levels are at 1, 2, 3, 5, 7, 9, 11, 13, 15,
17, 19 and 21~K~kms$^{-1}$.  {\bf c)} $^{12}$CO~J=2-1 integrated
intensity map of NGC~5775 obtained at the JCMT with a beam size of
21\arcsec. The map peak is 40.0~K~kms$^{-1}$. Contour levels are at 1,
2, 5, 8, 10, 15, 20, 25, 30 and 35~K~kms$^{-1}$.\\

Fig. 5: ROSAT X-ray image of NGC~5775 and its environment superimposed
on an optical image. \\

Fig. 6: {\bf a)} 23\arcsec-resolution 20~cm radio continuum contours
superimposed on the natural weighting HI total intensity
image. Contours are as in Fig.~1b.  {\bf b)} IRAS
12~$\mu$m contours (as in Fig.~3a) superimposed on the
natural weighting HI total intensity image.  {\bf c)} IRAS 25~$\mu$m
contours (as in Fig.~3b) superimposed on the natural
weighting HI total intensity image.  \\

Fig. 7: {\bf a)} H$\alpha$ contours superimposed on the R-band image.
Contours are at 70, 120 and 200 times 6.6 $\times$ 10$^{-18}$ ergs
s$^{-1}$ cm$^{-2}$ arcsec$^{-2}$.  The features labelled A, B, and C
are the brightest H$\alpha$ emission in the disk .  {\bf b)}
5\arcsec-resolution 20 cm radio continuum contours superimposed on the
H$\alpha$ greyscale.  {\bf c)} Radio continuum spectral index
(15\arcsec-resolution) superimposed on H$\alpha$.  {\bf d)}
Uniformly-weighted HI total intensity (13\arcsec-resolution) over
H$\alpha$.  {\bf e)} $^{12}$CO~J=2-1 integrated intensity
(21\arcsec-resolution) over H$\alpha$.\\

Fig. 8: The HI loop, F1, shown in comparison to data at other
wavebands.  In each frame, the HI has been integrated over velocities,
1683.4 to 1850.2 km s$^{-1}$ and the vertical extent of each frame is
119\arcsec.  {\bf a)} Integrated intensity HI.  The greyscale ranges
from 15 (white) to 200 (black) mJy beam$^{-1}$ km s$^{-1}$ and
contours are at 40, 50, 72, 100, 125, and 175 mJy beam$^{-1}$ km
s$^{-1}$, and so throughout unless otherwise indicated.  {\bf b)}
Intensity-weighted mean velocities over HI (here, the greyscale peak
flux is 180 mJy beam$^{-1}$).  Velocity contours are separated by 15
km s$^{-1}$.  {\bf c)} HI over H$\alpha$.  The greyscale ranges from
16 $\times$ 10$^{-19}$ \ergpers cm$^{-2}$ pixel$^{-1}$ (white,
2$\sigma$) to 64 $\times$ 10$^{-19}$ \ergpers cm$^{-2}$ pixel$^{-1}$
(black).  Two residual star images are visible in the H$\alpha$ image,
one at the center of the HI loop and one to the upper right.  {\bf d)}
6 cm radio continuum over HI.  Contours are 0.02, 0.03, 0.04, 0.06,
0.09, 0.15, 0.35 and 0.75 mJy beam$^{-1}$.  {\bf e)} 20 cm radio
continuum over HI.  Contours are 0.054, 0.075, 0.12, 0.18, 0.30, 0.50,
1.0, 2.0, and 3.5 mJy beam$^{-1}$.  {\bf f)} HI over spectral index
map.  The greyscale ranges from 0.8 (white) to 1.2 (black).  {\bf g
$\to$ i)} Position-velocity slices through the HI cube, taken along
vertical lines denoted {\it Left, Center,} and {\it Right} in (a).
Contours are at $\pm$0.6 (1.5$\sigma$), $\pm$0.8, $\pm$1.2, 1.8, 2.8,
5.0, 7.0, and 10 mJy beam$^{-1}$. \\

Fig. 9: The HI loop, F2, shown in comparison to data at other
wavebands.  In each frame, the HI has been integrated over velocities,
1683.4 to 1892.0 km s$^{-1}$ and the vertical extent of each frame is
147\arcsec.  {\bf a)} Integrated intensity HI.  The greyscale ranges
from 15 (white) to 300 (black) mJy beam$^{-1}$ km s$^{-1}$ and
contours are at 40, 50, 70, 100, 126, 156, 170, 200, and 300 mJy
beam$^{-1}$ km s$^{-1}$, and so throughout, unless otherwise
indicated.  {\bf b)} Intensity-weighted mean velocities over HI (here,
the greyscale peak is 250 mJy beam$^{-1}$ km s$^{-1}$).  Velocity
contours are separated by 15 km s$^{-1}$.  {\bf c)} HI over H$\alpha$.
The greyscale ranges from 16 $\times$ 10$^{-19}$ \ergpers cm$^{-2}$
pixel$^{-1}$ (white, 2$\sigma$) to 80 $\times$ 10$^{-19}$ \ergpers
cm$^{-2}$ pixel$^{-1}$ (black).  {\bf d)} 6 cm radio continuum over HI
(here the peak greyscale is 250 mJy beam$^{-1}$ km s$^{-1}$).
Contours are 0.04, 0.05, 0.065, 0.08, 0.15, 0.30 and 0.50 mJy
beam$^{-1}$. The hatched region indicates declining contours.  {\bf
e)} 20 cm radio continuum over HI (peak greyscale is 250 mJy
beam$^{-1}$ km s$^{-1}$).  Contours are 0.05, 0.07, 0.10, 0.15, 0.20,
0.30, 0.50, 1.0, and 2.0 mJy beam$^{-1}$.  {\bf f)} HI over spectral
index map.  The greyscale ranges from 0.8 (white) to 1.0 (black).
{\bf g) $\to$ i)} Position-velocity slices through the HI cube, taken
along vertical lines denoted {\it Left, Center,} and {\it Right} in
(a).  Contours are at $\pm$0.6 (1.5$\sigma$), $\pm$0.8, $\pm$1.2, 1.8,
2.8, 5.0, 7.0, and 10 mJy beam$^{-1}$.\\

Fig. 10: The HI loop, F3, shown in comparison to data at other
wavebands.  In each frame, the HI has been integrated over velocities,
1475.0 to 1683.4 km s$^{-1}$ and the vertical extent of each frame is
161\arcsec.  {\bf a)} Integrated intensity HI.  The greyscale ranges
from 15 (white) to 300 (black) mJy beam$^{-1}$ km s$^{-1}$ and
contours are at 40, 50, 70, 100, 125, 175, and 230 mJy beam$^{-1}$ km
s$^{-1}$, and so throughout, unless otherwise indicated.  {\bf b)}
Intensity-weighted mean velocities over HI (here, the peak greyscale
is 150 mJy beam$^{-1}$).  Velocity contours are separated by 15 km
s$^{-1}$.  {\bf c)} HI over H$\alpha$.  The greyscale ranges from 28
$\times$ 10$^{-19}$ \ergpers cm$^{-2}$ pixel$^{-1}$ (white,
3.5$\sigma$) to 64 $\times$ 10$^{-19}$ \ergpers cm$^{-2}$ pixel$^{-1}$
(black).  Two residual star images are visible in the H$\alpha$ image,
one to the upper left of the HI loop and one to the upper right.  {\bf
d)} 6cm radio continuum over HI (here, the peak greyscale is set to
200 mJy beam$^{-1}$ km s$^{-1}$).  Contours are 0.02, 0.03, 0.04,
0.06, 0.072, 0.09, 0.15, 0.30 mJy beam$^{-1}$.  {\bf e)} 20 cm radio
continuum over HI total intensity.  Contours are 0.050, 0.070, 0.10,
0.15, 0.20, 0.30, 0.50, and 1.0 mJy beam$^{-1}$.  {\bf f)} HI over
spectral index map.  The greyscale ranges from 0.85 (white) to 1.3
(black).  {\bf g) $\to$ i)} Position-velocity slices through the HI
cube, taken along vertical lines denoted {\it Left, Center,} and {\it
Right} in (a).  Contours are at $\pm$0.6 (1.5$\sigma$), $\pm$0.8,
$\pm$1.2, 1.8, 2.8, 5.0, 7.0, and 10 mJy beam$^{-1}$. \\

Fig. 11: Vertical position-velocity slice through the center of the HI
loop, F3, at the same position as in Fig.~10h, but including
the region below the major axis.  \\

Fig. 12: Sketch of an HI supershell, as typified by the features, F1,
F2, and F3 in NGC~5775. 

\end{document}